\documentclass[aps,preprint,11pt]{revtex4} 

\newcommand{\Sc}{{\cal S}}
\newcommand{\Ec}{{\cal E}}
\newcommand{\Rc}{{\cal R}}
\newcommand{\Uc}{{\cal U}}
\newcommand{\Mc}{{\cal M}}
\newcommand{\Lc}{{\cal L}}
\newcommand{\la}{\lambda}
\newcommand{\sv}{{\vec\sigma}}
\newcommand{\tr}{{\rm tr}}

\begin{document}
%
\title{Progressive Decoherence and Total Environmental 
Disentanglement} 
%
%
%
%
%
\author{Lajos Di\'osi}
%
%
%
\email{diosi@rmki.kfki.hu}\homepage{www.rmki.kfki.hu/~diosi}
\affiliation{Research Institute for Particle and Nuclear Physics\\
           H-1525 Budapest 114, P.O.Box 49, Hungary}


\begin{abstract}
The simple stationary decoherence of a two-state quantum system
is discussed from a new viewpoint of environmental entanglement.
My work emphasizes that an unconditional local state must totally be 
disentangled from the rest of the universe. It has been known for long 
that the loss of coherence within the given local system is gradual. 
Also the quantum correlations between the local system and the rest of 
the universe are being destroyed gradually. I show that, differently from 
local decoherence, the process of environmental disentanglement may 
terminate in finite time. The time of perfect disentanglement turns out to 
be on the decoherence time scale, and in a simple case we determine the 
exact value of it. 
\end{abstract}

\maketitle 

\section{Introduction}
Perfect isolation of a real physical system $\Sc$ is not possible. 
Even the simplest quantum system, like the spin of an electron, may be
in weak though continuous interaction with the environment $\Ec$. The  
sensitivity of quantum systems to environmental interactions goes
beyond the sensitivity of classical systems. In our example, the
coherence of local superpositions will asymptotically be lost at a certain 
decoherence time scale $\tau$. We consider the simplest isotropic model of 
environmental decoherence of a Pauli-spin vector $\sv$:
\begin{equation}\label{dc}  
\frac{d\rho}{dt}=-\frac{1}{4\tau}[\sv,[\sv,\rho]]~,
\end{equation}
where $\rho$ is the $2\times2$ density matrix of the spin.
This equation may, e.g., correspond to the proper statistical average 
\cite{Goretal78} of the Schr\"odinger-equation 
\begin{equation}\label{Sch}
\frac{d\psi}{dt}=-\frac{i}{2}{\vec\omega}\sv\psi
\end{equation} 
over the random magnetic field $\vec\omega$ which is $\tau^{-1/2}$-times
the standard isotropic white-noise.  
The solution $\rho(t)$ encodes the statistics of all possible local 
phenomena, i.e., the physics of the spin. It is, nonetheless,
known from the famous work of John Bell \cite{Bel66} that quantum 
correlations with remote systems paralyze the standard statistical 
interpretation of the local state $\rho$. We are going to analyze 
the fate of such external quantum correlations of our local system
during its irreversible evolution (\ref{dc}). We recapitulate
the definition of quantum correlation (entanglement) through
the notion of separability \cite{Wer89}. We can prove that all external
quantum correlations will disappear if and only if
\begin{equation}\label{log3}
t~~\geq~~\tau{\rm log}3~.
\end{equation}
The result comes from a recent theorem on `entanglement breaking
quantum channels' \cite{Sho02,Hol99,Rus02}. I will present an elementary
alternative proof. We shall mention a contrary case:
\begin{equation}\label{mc}
\frac{d\rho}{dt}=-\frac{1}{\tau}[\sigma_3,[\sigma_3,\rho]]~.
\end{equation}
This is continuous measurement of the $\sigma_3$ spin-component where
the total damping of external quantum correlations takes an infinite long 
time.
 
\section{Separability, entanglement, statistical consistency}
To study the structure of the external correlations of our system, 
we divide the universe $\Uc$ \cite{Uni} into
three parts: the local system $\Sc$ of interest, its environment $\Ec$ 
responsible for the irreversible evolution (\ref{dc}), and the rest $\Rc$ 
which has no current interaction with $\Sc$ but might have had it in the 
remote past. This time we concentrate on the correlations between $\Sc$ 
and $\Rc$. We do not consider $\Ec$ a dynamical system at all. We 
model it merely as a source of classical magnetic noise leading to the 
irreversible equation (\ref{dc}) for the local system $\Sc$. One says that 
the local system is \emph{classically correlated} with the rest of the 
universe if the total quantum state is \emph{separable} \cite{Wer89}:
\begin{equation}\label{sep}
\rho_\Uc=\sum_\la w^\la \rho^\la\otimes\rho_\Rc^\la~,
\end{equation}
i.e., it is the weighted mixture of uncorrelated (product) states.

Such an expansion is always possible for correlated classical systems.
Their density matrices are always diagonal. For the corresponding
classical densities the separability condition (\ref{sep}) reads:
\begin{equation}\label{sepcl}
\rho_\Uc(x,X)=\sum_\la w^\la \rho^\la(x)\rho_\Rc^\la(X)~,
\end{equation}
where $x$ and $X$ label the states of the classical systems $\Sc$ and
$\Rc$, respectively. To construct the r.h.s., first we identify $\la$ by the 
composite label $(x',X')$ and we identify $w^\la$ by $\rho_\Uc(x',X')$. Then 
we can make the choices \mbox{$\rho^\la(x)=\delta(x-x')$} and 
\mbox{$\rho_\Rc^\la(X)=\delta(X-X')$}. 
  
Quantum systems differ radically. They may not satisfy the separability 
(\ref{sep}). By definition, the local system $\Sc$ is \emph{quantum-correlated
(entangled)} with the rest $\Rc$ of the universe if the composite state 
$\rho_\Uc$ is non-separable. Most typically, this is the case when the 
universe is in a pure state $\rho_\Uc=\vert\Uc\rangle\langle\Uc\vert$ which 
is not a product state:
\begin{equation}\label{Uent}
\vert\Uc\rangle\neq\vert\Sc\rangle\otimes\vert\Rc\rangle~.
\end{equation}

Why do any sort of correlation with the remote part of the universe should 
bother our observations on the local system? The reason is tricky. If the 
separability condition (\ref{sep}) holds then the correlation between local 
phenomena in $\Sc$ and remote phenomena in $\Rc$ can in principle be 
explained on the ground of classical statistics \cite{Wer89}. If, however,
the separability condition does not hold then the local and remote 
phenomena may become inconsistent from the viewpoint of classical 
statistical rules unless we allow instantaneous signal propagation between 
the local and remote systems $\Sc$ and $\Uc$, respectively \cite{Bel66}. 
Such faster-than-light propagation is impossible because it contradicts to
our notion of locality. The external quantum-correlations (entanglements) of 
a local system will thus object that the usual statistical interpretation 
of the local system be consistent with the simultaneous statistical 
interpretation of other possible systems populating the rest of the universe. 
\emph{It is therefore not possible to claim that $\rho$ is the unconditional 
local state unless we make sure that the local system $\Sc$ is totally 
disentangled from the rest $\Rc$ of the universe}. There should be no
quantum correlations left between them.

\section{Total disentanglement}
We shall discuss the existence and fate of the external quantum
correlations of our chosen system.
The formal solution of the equation (\ref{dc}) contains a linear
time-dependent map $\Mc(t)$: 
\begin{equation}\label{sol}
\rho(t)=\Mc(t)\rho(0)~.
\end{equation}
The state $\rho(t)$ does obviously not encode external correlations, i.e., 
those between the two-state system $\Sc$ and the rest of the universe $\Rc$.   
The generic state $\rho_\Uc$ does not satisfy the separability condition 
(\ref{sep}) because usually it contains quantum correlations between $\Sc$ 
and $\Rc$. So does the initial state $\rho_\Uc(0)$. Surprisingly, a recent 
theorem \cite{Sho02,Hol99,Rus02} enables us to conclude that after a finite 
threshold time the state $\rho_\Uc(t)$ becomes totally separable, i.e., the 
local system $\Sc$ disentangles from the rest $\Rc$ of the universe. The 
power of the theorem lies in that it needs only the local map $\Mc(t)$ 
though with the tacit assumption that $\Sc$ and $\Rc$ evolve independently. 

Accordingly, a certain map $\Mc$ will completely disentangle the local 
system if and only if it acts as follows: 
\begin{eqnarray}\label{theo}
&&\Mc\rho=\sum_\la p^\la\rho^\la~,\nonumber\\
&&p^\la=\tr[P^\la\rho]~,
\end{eqnarray}
where $\{\rho^\la\}$ is a certain set of states that does not depend on the 
original state $\rho$. The mixing probabilities $\{p^\la\}$ depend on 
$\rho$ via a generalized measurement, i.e., the set $\{P^\la\}$ must be a 
positive-operator-valued-measure (POVM) \cite{JauPir67}:
\begin{equation}
P^\la\geq0~,~~~\sum_\la P^\la=I~.
\end{equation}

One confirms easily that the above conditions are sufficient for the 
separability of the mapped composite state. Recall that we assumed 
independent maps for $\Sc$ and $\Rc$: 
\begin{equation}
\Mc_\Uc\rho_\Uc=\left(\Mc\otimes\Mc_\Rc\right)\rho_\Uc~.
\end{equation}
By inserting $\Mc$ from Eq.~(\ref{theo}), where we set
\begin{eqnarray}
&&p^\la=\tr\left[(P^\la\otimes\Mc_\Rc)\rho_\Uc\right]~,\nonumber\\
&&\rho^\la_\Rc
     =\frac{1}{p^\la}\tr_\Sc\left[(P^\la\otimes\Mc_\Rc)\rho_\Uc\right]~,
\end{eqnarray}
the separable form (\ref{sep}) is obtained for $\Mc_\Uc\rho_\Uc$.
The proof of necessity of the condition (\ref{theo}) is a harder task.

\section{The disentanglement time} 
Let us test the condition (\ref{theo}) on the map $\Mc(t)$ solving the 
Eq.~(\ref{dc}) for $t\geq0$. The identical map $\Mc(0)$ can of course never 
admit the form (\ref{theo}). Neither does $\Mc(t)$ at short times. 
But a threshold time will exist after which the map $\Mc(t)$ can already be 
written into the desired form (\ref{theo}). The irreversible equation 
(\ref{dc}) describes an exponential isotropic depolarization. Its solution 
(\ref{sol}) can be written into this concrete form:
\begin{equation}\label{sol1}
\rho(t)\equiv\Mc(t)\rho(0)=\frac{1}{2}
\left[I+e^{-t/\tau}\sv\tr[\sv\rho(0)]\right]~.
\end{equation}
If follows from a theorem in Ref.~\cite{Rus02} that the map $\Mc(t)$
is entanglement breaking if and only if $3e^{-t/\tau}\leq1$ which is
the condition (\ref{log3}). Here I present a constructive proof. 
In order to put the time-dependent map $\Mc(t)$ into the desired form 
(\ref{theo}), let us try the following. We assume the set $\{\rho^\la\}$ 
to consist of six states labeled by $\la=(\alpha,s)$: 
\begin{equation}\label{states}
\rho^{\alpha s}
=\frac{1}{2}\left[I+3se^{-t/\tau}\sigma_\alpha\right]~,
\end{equation}
where $\alpha=1,2,3$ while $s=\pm1$ in turn. Let the corresponding six POVM 
elements be proportional to six different completely polarized pure state 
projectors:
\begin{equation}\label{POVM}
P^{\alpha s}
=\frac{1}{6}\left[I+s\sigma_\alpha\right]~.
\end{equation}
By inserting the Eqs.~(\ref{states},\ref{POVM}) into (\ref{theo}),
we obtain the solution (\ref{sol1}) indeed. However, a closer look at 
the Eq.~(\ref{states}) warns us that the states $\rho^{\alpha s}$ do not 
exist until the absolute value of the coefficient $3e^{-t/\tau}$ descends
from the initial value $3$ to $1$. This completes our proof.  The
map $\Mc(t)$ disentangles any initial state $\rho(0)$ from the rest
of the universe after time (\ref{log3}). 

Let us assume that $\Rc$ is just another two-state system and its evolution 
is trivial: it does not evolve at all. Then the composite state satisfies 
the trivial extension of the local evolution equation (\ref{dc}):
\begin{equation}
\frac{d\rho_\Uc}{dt}=
-\frac{1}{4\tau}[\sv\otimes I_\Rc,[\sv\otimes I_\Rc,\rho_\Uc]]~.
\end{equation} 
Assuming the initial composite state $\rho_\Uc(0)$ is the maximally 
entangled singlet state, we can find the following time-dependent solution:
\begin{equation}
\rho_\Uc(t)
=\frac{1}{4}\left[I\otimes I_\Rc-e^{-t/\tau}\sv\otimes\sv\right]~.
\end{equation}
This state remains entangled for $e^{-t/\tau}>1/3$ (c.f. \cite{Wer89})
which shows that the threshold (\ref{log3}) can not be sharpened.

\section{Discussion and outlook}
I have analyzed the simplest case of noisy environment (\ref{dc}) which is 
capable to destroy all external quantum correlations of a local system in
finite time (\ref{log3}). It is important to note that classical correlations
may survive the death of quantum ones. The studied environment is classical 
and non-dynamical, its simplest realization is an isotropic classical magnetic
white-noise field. The corresponding map is called the `depolarizing chanel' 
in quantum communication theory \cite{NieChu00}. There is a non-isotropic 
environment producing decoherence on the same time scale as Eq.~(\ref{dc}) did
but with a different structure (\ref{mc}). This is called the `measurement 
channel' since for infinite time it describes the ideal measurement of the 
spin-component $\sigma_3$. Obviously it disentangles the local system $\Sc$ 
from $\Rc$ at time $t=\infty$ as any ideal spin-measurement should. Regarding 
finite times, however, the external quantum correlations may always survive.
When I tried the sort of constructive proof, successful for the
`depolarizing chanel' (\ref{dc}), it failed for the `measurement one' 
(\ref{mc}). An ultimate proof may be learned in Ref.~\cite{Rus02}: the 
`measurement chanel' only eliminates external quantum correlations at 
$t=\infty$, i.e., not before the ideal measurement is completed.
Recently, I have found a third, affirmative, example. From earlier results of 
Kiefer and myself \cite{DioKie02} on the emergent exact positivity of
the Wigner- and the P-functions, it follows that a free particle in simplest 
position decohering environment will become disentangled in finite time. 
This result may certainly be extended for the broader class \cite{Dek81} of 
evolution equations including harmonic potential and friction as well.

For future investigations, the generic problem can be formulated in the
following way. Suppose the environment induces a semi-group evolution for
the local system, described by the generator $\Lc$ \cite{Lin76Goretal76}:
\begin{equation}
\frac{d\rho}{dt}=\Lc\rho~. 
\end{equation}
We construct the time-dependent map of the semi-group: 
\begin{equation}
\Mc(t)=\exp(\Lc t)~,
\end{equation}
which solves the evolution equation. Then one has to apply the condition 
(\ref{theo}) to the above map in order to determine if it disentangles in
finite/infinite time or whether it disentangles at all. 

\section*{Acknowledgment(s)}
I am grateful to the organizers and sponsors of the conference on 
"Irreversible Quantum Dynamics" for the invitation to talk in Triest and
to contribute to this volume. My research was also supported by the 
Hungarian OTKA Grant No. 032640.

%

\end{document}